# Tunable Strong Magnon-Magnon Coupling in Two-Dimensional Array of Diamond Shaped Ferromagnetic Nanodots

Sudip Majumder[1], Samiran Choudhury[1], Saswati Barman[2], Yoshichika Otani[3,4], Anjan Barman[1,*]

[1]Department of Condensed Matter Physics and Material Sciences, S. N. Bose National Centre for Basic Sciences, Block JD, Sector III, Salt Lake, 700106, Kolkata, India
[2]Institute for Engineering and Management, Sector V, Salt Lake, 700091, Kolkata, India
[3] CEMS-RIKEN, 2-1 Hirosawa, Saitama, 3510198, Wako, Japan
[4]Institute for Solid State Physics, University of Tokyo, 515 Kashiwanoha, Chiba, 277 8581, Kashiwa, Japan
*Email: abarman@bose.res.in

## Abstract

Hybrid magnonics involving coupling between magnons and different quantum particles have been extensively studied during past few years for varied interests including quantum electrodynamics. In such systems, magnons in magnetic materials with high spin density are utilized where the "coupling strength" is collectively enhanced by the square root of the number of spins to overcome the weaker coupling between individual spins and the microwave field. However, achievement of strong magnon-magnon coupling in a confined nanomagnets would be essential for on-chip integration of such hybrid systems. Here, through intensive study of interaction between different magnon modes in a $Ni_{80}Fe_{20}$ (Py) nanodot array, we demonstrate that the intermodal coupling can approach the strong coupling regime with coupling strength up to 0.82 GHz and cooperativity of 2.51. Micromagnetic simulations reveal that the intermodal coupling is mediated by the exchange field inside each nanodot. The coupling strength could be continuously tuned by varying the bias field ($H_{ext}$) strength and orientation ($\phi$), opening routes for external control over hybrid magnonic systems. These findings could greatly enrich the rapidly evolving field of quantum magnonics.

## 1. Introduction

Hybrid quantum systems [1] have recently attracted great attention due to their fundamental importance and potential applications. It provides a new paradigm for the coherent transfer of quantum states from one platform to another to execute quantum information processing [2,3]. This significantly facilitates the research on the fundamental physics of coupling between different platforms which may lead to varied applications of quantum technologies, such as: quantum computing [4,5], quantum communications [6,7], and quantum sensing [8]. The introduction of magnons in hybrid systems was initiated from the exploration of spin ensembles coupled to microwave photons [8-10]. The higher densities of spin in magnetic materials and their collective dynamics as magnons, provide ultra-strong coupling with cooperativity up to $10^3$-$10^4$ [11,12]. During the last decade, extensive research has been done on magnon-magnon coupling [13-19]. However, on-chip integration of hybrid systems requires downscaling the dimensions of the systems to the nanometer range. The microwave cavity usually has the dimension of millimeters. The coupling strength ($g$) is proportional to the square root of the number of spins present in the magnetic material [20,21]. To increase the coupling strength the number of spins in the magnetic material is usually required to be large enough (N > $10^{13}$), thereby restricting the size of the microwave cavity and magnet and the ensuing device miniaturization towards CMOS integration.

To overcome this geometrical limitation of a microwave cavity, it becomes imperative to search for different systems to act as nanometric resonators. In this context, the recent development of interlayer magnon coupling or exchange-driven magnon-magnon coupling in the magnetic systems has opened a new avenue for quantum magnonics [22-24]. In the last decade, extensive studies have been done using both confined and propagating magnons in the field of magnonics, which emerged as an exciting field of research. To this end single nanomagnets have been studied extensively due to their geometrically confined rich volume and localized magnetic modes [25-29] in nanometer dimension and their tunability with different external parameters. Therefore, such systems possess great potential in quantum magnonics with the possibility of developing magnon-based on-chip quantum information processing systems in the GHz and THz frequency range with high energy efficiency. Recently magnon-magnon coupling has been observed experimentally in ferromagnetic nanowire array[15] and in single nanomagnet using micromagnetic simulation[30]. Furthermore, moderate to strong magnon-magnon coupling have also been observed in $Ni_{80}Fe_{20}$ (Py) nanocross array mediated by dynamic dipolar interaction [31] and anisotropic dipolar interaction[32]. These studies have opened a new approach for executing and controlling this phenomenon in a large variety of systems by tailoring the geometric and material parameters of these artificially patterned systems and the external bias field. This leads the quest for optimal solutions for applications in magnon-based quantum information technology.

Here, we have explored magon-magnon coupling in diamond-shaped Py nanodot array with the aid of a broadband ferromagnetic resonance (FMR) spectrometer[33,34] and micromagnetic simulations. Remarkably, we observe an avoided crossing (anticrossing) of magnon modes [1] characteristic of the formation of hybrid system. Anticrossing gap of up to 0.82 GHz and the ensuing cooperativity value as high as 2.51 are observed. Micromagnetic simulations reveal that the coupling between two magnon modes is mediated by the exchange field within each nanodot. Furthermore, the coupling strength is found to be highly dependent on the orientation and strength of the bias magnetic field, leading towards the possibility of externally controlled hybrid magnonic devices.

## 2. Experimental Details

The 20-nm-thick diamond shaped Py nanodots, arranged in an array of dimensions 25 μm × 200 μm, were prepared on self-oxidized Si [100] substrate by using electron beam evaporation (EBE), electron beam lithography (EBL), and $Ar^+$ ion milling tools. A coplanar waveguide (CPW) made of Au, having 150 nm thickness, 30 μm wide central conducting (signal) line and 50 Ω characteristic impedance (Fig. 1(a)) was deposited on top of each array for broadband FMR measurements. The CPW is separated from the nanodot array by a 60-nm-thick insulating $Al_2O_3$ layer. The fabrication details are described in section S1 of the Supplementary Materials. Fig. 1(b) exhibits the scanning electron microscope (SEM) image of the diamond nanodot array arranged in a square lattice having width and height of the nanodots as 325 nm ($d_x$) and 350 nm ($d_y$) and lattice constant of 400 nm. The nanomagnet's lateral dimensions and pitch are shown in the SEM image of Fig. 1(b). The SEM image shows that the fabricated structures suffer from slight edge deformations and rounded corners. All these deformations have been incorporated in the micromagnetic simulations as described later. The applied bias magnetic field orientation is shown in the inset of Fig. 1(b). The spin-wave (SW) spectra from the samples were measured using a broadband FMR spectrometer, consisting of a high-frequency Vector Network Analyzer (VNA, Agilent PNA-L, model no.: N5230C, frequency range: 10 MHz to 50 GHz) and a homemade high-frequency probe station equipped with nonmagnetic ground-signal-ground (GSG)-type picoprobe (GGB Industries, model no.: 40A-GSG-150-EDP) and a coaxial cable. One end of the CPW is shorted and the back-reflected signal is collected and fed back to the VNA by the same GSG probe and the coaxial cable. From the frequency dependent real part of the S-parameter in the reflection geometry (Re ($S_{11}$)), different SW frequencies are identified, which results in the characteristic SW spectrum of the sample. Additional details of the experimental setup are given in section S2 of the Supplementary Materials.

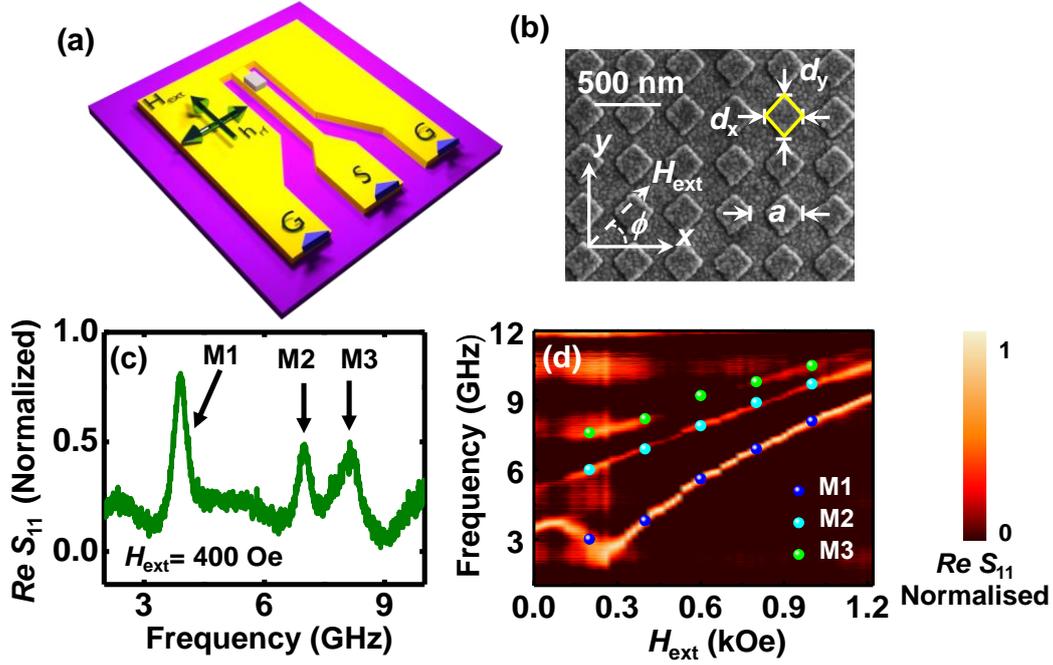

FIG. 1. (a) Schematic of the experimental geometry. The directions of the bias magnetic field ($H_{ext}$) and rf magnetic field ($h_{rf}$) are shown in the schematic. (b) SEM image of diamond-shaped $Ni_{80}Fe_{20}$ (Py) nanodots arranged in a square lattice having lattice constant $a$ = 400 nm and nanodot width $d_x$ = 325 nm, height $d_y$ = 350 nm. The inset again shows the orientation of $H_{ext}$ with respect to $h_{rf}$. (c) Real parts of the forward scattering parameter ($S_{11}$) representing the FMR spectra at $H_{ext}$ = 400 Oe applied at an azimuthal angle $\phi$ = 0°. The observed spin-wave (SW) modes are marked by down arrows. (d) Bias field ($H_{ext}$) dependent SW absorption spectra of Py nanodots is shown at $\phi$ = 0°. The surface plots correspond to the experimental results, while the symbols represent the simulated data. The color map for the surface plots and the schematic of $H_{ext}$ are given at the bottom right corner of the figure.

## 3. Results and Discussion

### 3.1. Experimental Result

#### 3.1.1. Field Dependence of SW

The SW absorption spectra (Re ($S_{11}$)) are acquired from FMR measurements for a broad range of bias magnetic field. Fig. 1(c) shows representative raw spectra at $H_{ext}$ = 400 Oe. At first, the magnetization of the samples are saturated along the +x direction by applying $H_{ext}$ = 1800 Oe, followed by gradual reduction of the field from 1600 Oe to 0 Oe at steps of 20 Oe in a single trace. The surface plot in Fig. 1(d) displays the bias-field-dependent of SW absorption spectra with their maximum power normalized to 1.0. These surface plots are generated from the individual Re ($S_{11}$) spectra acquired at a given applied magnetic field. Here, the bright regions represent the experimental data while the symbols represent the micromagnetic simulation results. The normalized surface plots help to identify three separate branches of SW, among which the lowest frequency branch M1 shows maximum intensity in the entire field regime. As we decrease the bias field M1 shows a dip (minimum) in $f$-$H_{ext}$ at $H_{ext}$ ≈ 300 Oe, which indicates a mode softening due to transition in magnetization state of the nanomagnet array. Other two SW modes M2 and M3 do not show any such transition and monotonically decrease with the reduction in the bias field.

Fig. 2 shows the magnetic field dependences of the frequencies at different bias field angles. The variation of magnetic field orientation creates some remarkable changes. First, the dip in M1 occurring at ~300 Oe gradually disappears. Fig. 2(a) shows the $f$-$H_{ext}$ plot at $\phi = 5°$, where the dip shows an upward shift. At $\phi = 15°$, the dip completely disappears and the M1 shows a monotonic variation of frequency with the field, as shown in Fig. 2(b). Secondly, the relative intensity of M2 and M3 shows a clear variation with the bias field orientation. For $5° \leq \phi \leq 15°$, M2 gradually losses its intensity at the expense of gradual increment of intensity of M3, which starts to dominate over M2 at $\phi = 15°$. With further increment of angle, M2 further loses its intensity and at $\phi = 23°$ it completely disappears. Fig. 2(c) shows the $f$-$H_{ext}$ plot at $\phi = 23°$ where a clear anticrossing between the branches representing modes M1 and M3 is observed at $H_{ext}$ = 1060 Oe. The vertical dotted line represents the anticrossing field ($H_{ac}$) in the $f$-$H_{ext}$ plot. The value of $H_{ac}$ gradually shifts towards the lower field regime as we keep increasing $\phi$.

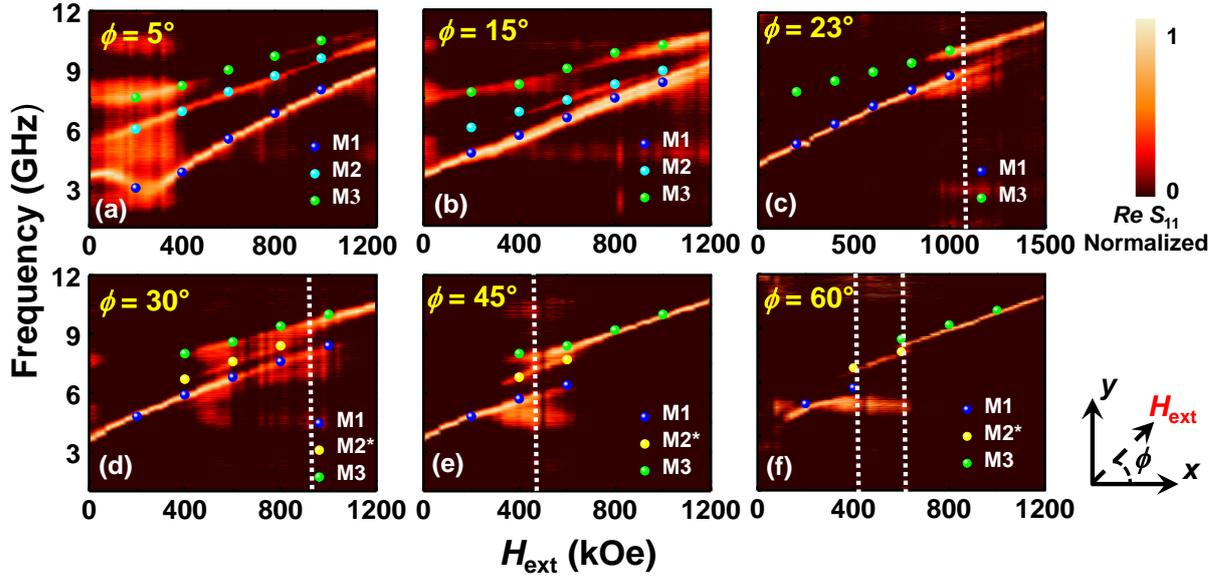

FIG. 2. Bias field ($H_{ext}$) dependent SW absorption plots of Py diamond shaped nanodot array are shown for the bias field orientation ($\phi$) of (a) 5°, (b) 15°, (c) 23°, (d) 30°, (e) 45° and (f) 60°. The surface plots correspond to the experimental results, while the symbols represent the simulated data. The color map for the surface plots and the schematic of the external applied field ($H_{ext}$) are given at the bottom right corner of the figure.

Fig. 2(d) shows the magnetic field dispersion of SW frequencies at $\phi = 30°$ where an anticrossing is observed at $H_{ext} = 920$ Oe in between the SW modes M1 and M3. Here, the mid frequency SW mode M2* reappears, though the intensity of this mode is low. With further increment of $\phi$, this mode becomes more prominent and two different anticrossings are now observed instead of one. One of those appears in between M1 and M2* and another one in between M2* and M3. At $\phi = 45°$, both of the anticrossings are observed at $H_{ext} = 475$ Oe as shown in Fig. 2(e). With further increment of $\phi$, the first anticrossing shifts towards lower bias magnetic field values, whereas the second one appears in higher bias field values. Fig. 2(f) shows the magnetic field dispersion of SW frequencies at $\phi = 60°$ where the first anticrossing in between M1 and M2* appear at $H_{ext} = 410$ Oe and second one at $H_{ext} = 600$ Oe.

### 3.1.2. Angular Dependence of SW

The variation of SW modes and their mutual interactions show high dependence on the in-plane magnetic field orientation $\phi$. For this reason, $\phi$-dependence of SW spectra were acquired at a constant bias field magnitude $H_{ext}$ in the range $0° \leq \phi \leq 360°$. In Fig. 3(a-d), we have presented the $\phi$-dependence at $H_{ext}$ = 200, 400, 600 and 800 Oe. To show the anticrossing points we have magnified the relevant regions of the $\phi$-dependent SW spectra. In the Supplementary Information figure S4, we have shown the full range of $\phi$-dependence. At a lower field value like $H_{ext}$ = 200 Oe, only M1 shows angular dispersion as shown in Fig. 3(a). With an increment in $H_{ext}$, two more modes start to show angular dispersion. Here, mode M1 shows a sharp variation of frequency with a minimum at $\phi = 0°$, corresponding to the minimum observed in Fig. 1(d). As we increase the field this sharp modulation gradually transforms into a continuous angular variation. Fig. 3(b) shows the angular dispersion at $H_{ext}$ = 400 Oe. For $\phi$ between 50° and 55°, an anticrossing gap appears in between M1 and M2* which is shown by a white dotted line. At a higher field of $H_{ext}$ = 600 Oe instead of one, two different anticrossings are observed. The first one appears in between M1 and M3 at $\phi = 40°$ while the 2nd one appears in between M2* and M3 at $\phi = 60°$. With an increment of magnetic field (e.g., 800 Oe) the first anticrossing shifts towards lower angle (e.g. 35°), while the second one gradually disappears as shown in Fig. 3(d). Due to four fold symmetry[35] of diamond shaped nanodot array these anticrossing also appear in other three quadrants of angular variation spectra of SW, which is shown in section S4 of supplementary section.

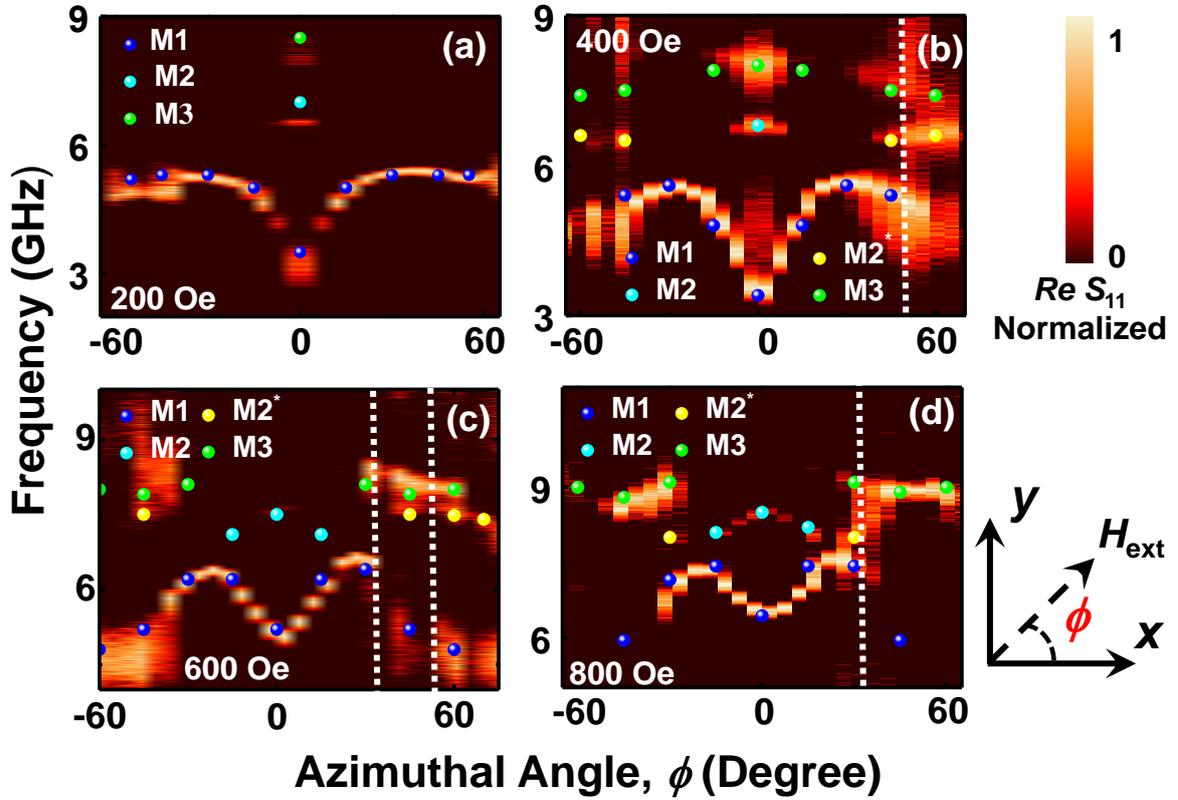

FIG. 3. Variation of SW frequency as a function of the azimuthal angle ($\phi$) varying from 0° to 360° for bias field value fixed at (a) $H_{ext}$ = 200 Oe, (b) 400 Oe, (c) 600 Oe and (d) 800 Oe. The surface plots correspond to the experimental results, while the symbols represent the simulated data. The colour map for the surface plots and the schematic of $H_{ext}$ are shown on the right side of the figure.

### 3.1.3. Anticrossing Strength

Fig. 4(a) shows the power spectrum measured at $H_{ext}$ = 1060 Oe, which is the anticrossing field ($H_{ac}$) for $\phi = 23°$ configuration. The blue line represents the FMR spectra whereas the red line represent the fitted spectra using an antisymmetric lorentzian function. Other FMR spectra for varying anticrossing fields are presented in section S5 of Supplementary Information. The magnon–magnon coupling strength $g$ is defined as half of the peak-to-peak frequency spacing at the anticrossing field, which is shown in Fig. 4(a). In order to estimate the strength of interaction between these two modes, we have extracted the value of $g_{13}$ and the corresponding dissipation rates $\kappa_1$, $\kappa_3$ as shown in Fig. 4(a). Here, $\kappa_1$ and $\kappa_3$ are defined as half-width at half-maximum of the FMR peak of SW mode M1 and M3, respectively.

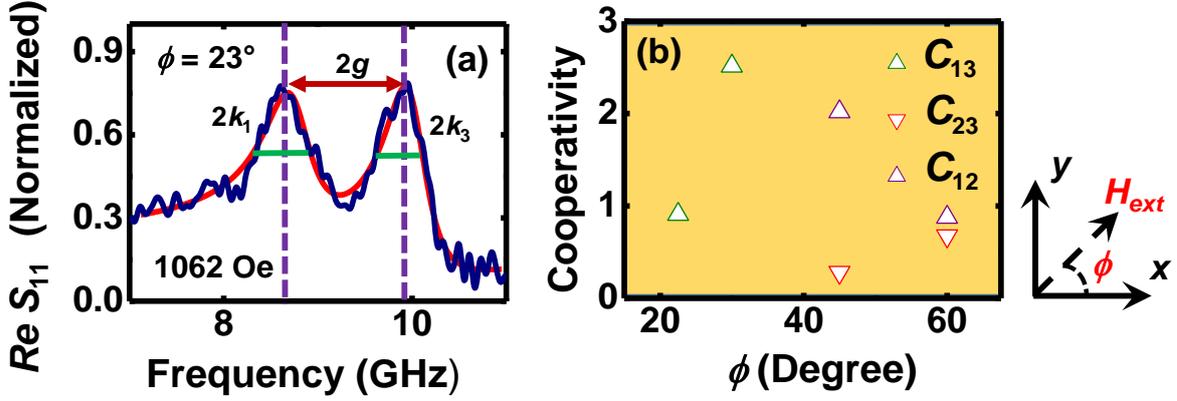

FIG. 4. Real part of $S_{11}$ parameter as a function of frequency to highlight the anticrossing field are shown for $\phi =$ (a) 23°. The frequency gap in the anticrossing mode reveals the coupling strength $g$. (b) Variation of cooperativity factor with the orientation of bias field. It shows that coupling strength is stronger at $\phi = 30°$ and 45°. The schematic of $H_{ext}$ are shown on the right side of the figure.

|  | $g_{13}$ (GHz) | $g_{12}$ (GHz) | $g_{23}$ (GHz) | $\kappa_1$(GHz) | $\kappa_2$(GHz) | $\kappa_3$(GHz) | $C_{13}$ | $C_{12}$ | $C_{23}$ |
|---|---|---|---|---|---|---|---|---|---|
| 23° | 0.592 | - | - | 0.60 | - | 0.711 | 0.821 |  |  |
| 30° | 0.82 | - | - | 0.423 | - | 0.660 | 2.515 |  |  |
| 45° | - | 0.745 | 0.255 | 0.426 | 0.645 | 0.645 | - | 2.019 | 0.113 |
| 60° | - | 0.915 | 0.205 | 1.35 | 0.69 | 0.707 | - | 0.878 | 0.675 |

**Table 1** The extracted values of coupling strength ($g$), FWHM ($2k$) and calculated cooperativity factor ($C$) for different orientation of bias field at the anticrossing points. Values of $g$ and $k$ are extracted from the FMR spectra).

At $\phi = 23°$ the extracted value of $g_{13}$ is 0.592 GHz, while the values of $\kappa_1$ and $\kappa_3$ are found to be 0.60 GHz and 0.711 GHz, respectively. Since $g_{13} < \kappa_1$ and $\kappa_3$, therefore the interaction between M1 and M3 can be considered as weak coupling. In the opposite case, i.e. when $g_{13} > \kappa_1$ and $\kappa_3$ it will be considered as strong coupling between two SW branches. We have also calculated magnon–magnon cooperativity ($C$), which is defined as $C_{\alpha\beta} = g^2/(\kappa_\alpha \kappa_\beta)$ ($\alpha,\beta = 1, 2, 3$) and obtained $C_{13} = 0.821$ for the coupling between M1 and M3. The extracted value of $g_{\alpha\beta}$, $k_\alpha$, $k_\beta$, and the estimated value of $C_{\alpha\beta}$ for anticrossing points corresponds to different bias field angles are listed in Table 1. At $\phi = 30°$ obtained value for $g_{13}$, $\kappa_1$, $\kappa_3$ and $C_{13}$ are estimated 0.82, 0.423, 0.66, and 2.515, respectively and here this magnon-magnon coupling falls in the strong coupling regime. From Table 1, we can see that first anticrossing at $\phi = 45°$ also shows strong magnon-magnon coupling with $C = 2.019$, while the second one shows weak interaction. At $\phi = 60°$ both the interactions are in the weak coupling regime. Fig. 4(b) shows the $\phi$-dependence of the $C$ where it shows the tunability of coupling strength with the in-plane magnetic field orientation. It also exhibits that the interaction between different SW branches show strong coupling in-between 30° to 45° orientation.

### 3.2. Micromagnetic Simulation

#### 3.2.1. Static Magnetic Configuration

In Fig. 1(d) at $\phi = 0°$, a sharp minimum is observed which gradually vanishes for higher values of $\phi$. The answer to this lies in the nanodot structure and its rich and flexible spin configurations which we have simulated using OOMMF software[36]. Details of the micromagnetic simulations are given in section S3 of the Supplementary Materials. The simulations reproduce important features of the experimental SW spectra with nearly identical frequencies and number of modes besides their relative intensity variations. The simulated static spin textures within the nanomagnet array for different bias field magnitudes $H_{ext}$ at $\phi = 0°$ and 45° are shown in Fig. 5. At $\phi = 0°$, the nanodot structure shows drastic variation in spin configurations with $H_{ext}$. It shows the formation of an S-state at the lower field regime ($H_{ext} = 100$ Oe) as shown in Fig. 5. At larger bias fields (e.g., $H_{ext} = 800$ Oe), the spins are nearly aligned along the bias-field direction (x-axis) and switch to a leaf-state (Fig. 5). This transformation from S- to leaf-state occurs for 250 Oe $\leq H_{ext} \leq$ 350 Oe, where the SW frequency shows a minimum as a function of $H_{ext}$. At $\phi = 45°$, this transformation is not observed. Here, for the entire field range, the static magnetic configuration shows a leaf state.

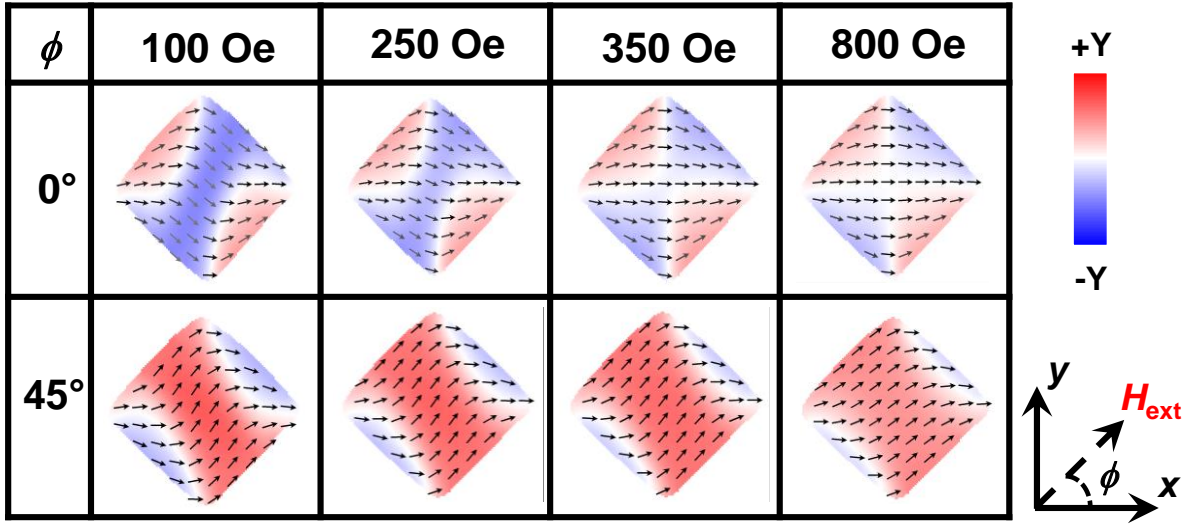

FIG. 5. Simulated static magnetic configurations for Py nanodot array at four different bias magnetic-field magnitude ($H_{ext}$) at $\phi = 0°$ and $\phi = 45°$. We have shown here a single nanodot from the center of the array for clarity in spin configurations. The nanodot structure shows a drastic variation in spin configurations with bias magnetic-field strength.

### 3.2.2. SW mode Characterization

To interpret the nature of the SW modes, we have further simulated the spatial profiles of power and phase of each SW mode by using a home-built MATLAB based code Dotmag[37]. OOMMF simulation provides magnetization ($M(r, t)$) information of each rectangular prism-like cell at different simulation times. By performing discrete Fourier transformation with respect to time in each of these cells and subsequently extracting the power and phase of the dynamic magnetization for a desired frequency gives rise to the spatial distribution of the power phase profile for that particular mode. In Fig. 6, we have shown the power distribution profile of SW mode at $\phi = 45°$ orientation for five different fields, $H_{ext} = 200$ Oe ($H_{ext} << H_{ac}$), 400 Oe ($H_{ext} < H_{ac}$), 475 Oe ($H_{ac}$), 600 Oe ($H_{ext} > H_{ac}$) and 1000 Oe ($H_{ext} >> H_{ac}$), while the phase profile for each case is shown in the inset. The power profile at $H_{ext} = 1000$ Oe indicates that at high bias field only existing mode is M3, which is boosted by all the available energy. With a gradual decrement of bias field, two additional modes M1 and M2 appear and the power of M3 is gradually transferred to these two modes. At the anticrossing field, $H_{ext} = 475$ Oe, M2 appears as the most intense mode although M1 and M3 have significant power at this field. At lower fields, this power is gradually transferred to M1, and at 200 Oe, barring M1 other modes disappear. Similar to this energy exchange, the phase profiles also exhibit interchange of mode behavior. At high bias fields (e.g., 1000 Oe), M3 shows quantized nature in BV-like geometry with a quantization number $n = 3$. With a decrease in the field, this mode gradually transforms into higher-order quantized mode and M2* is transformed into a quantized mode with $n = 3$. At $H_{ext} = 475$ Oe the quantization number of M1, and M3 are $n = 5$, and 7, respectively, while for M2*, $n = 3$, which is identical to the quantization number of M3 at $H_{ext} = 1000$ Oe. This transformation of mode quantization number is also seen in-between M1 and M2* as we further reduce the bias field and finally at $H_{ext} = 200$ Oe, M1 shows a quantized behavior with $n = 3$. This transformation of power as well as mode property from one branch of SW mode to another at the anticrossing region indicates a strong interaction between these modes. For other orientation like $\phi = 23°$, 30° and 60°, similar kind of behavior are observed, which are shown in section S6 of the Supplementary Materials.

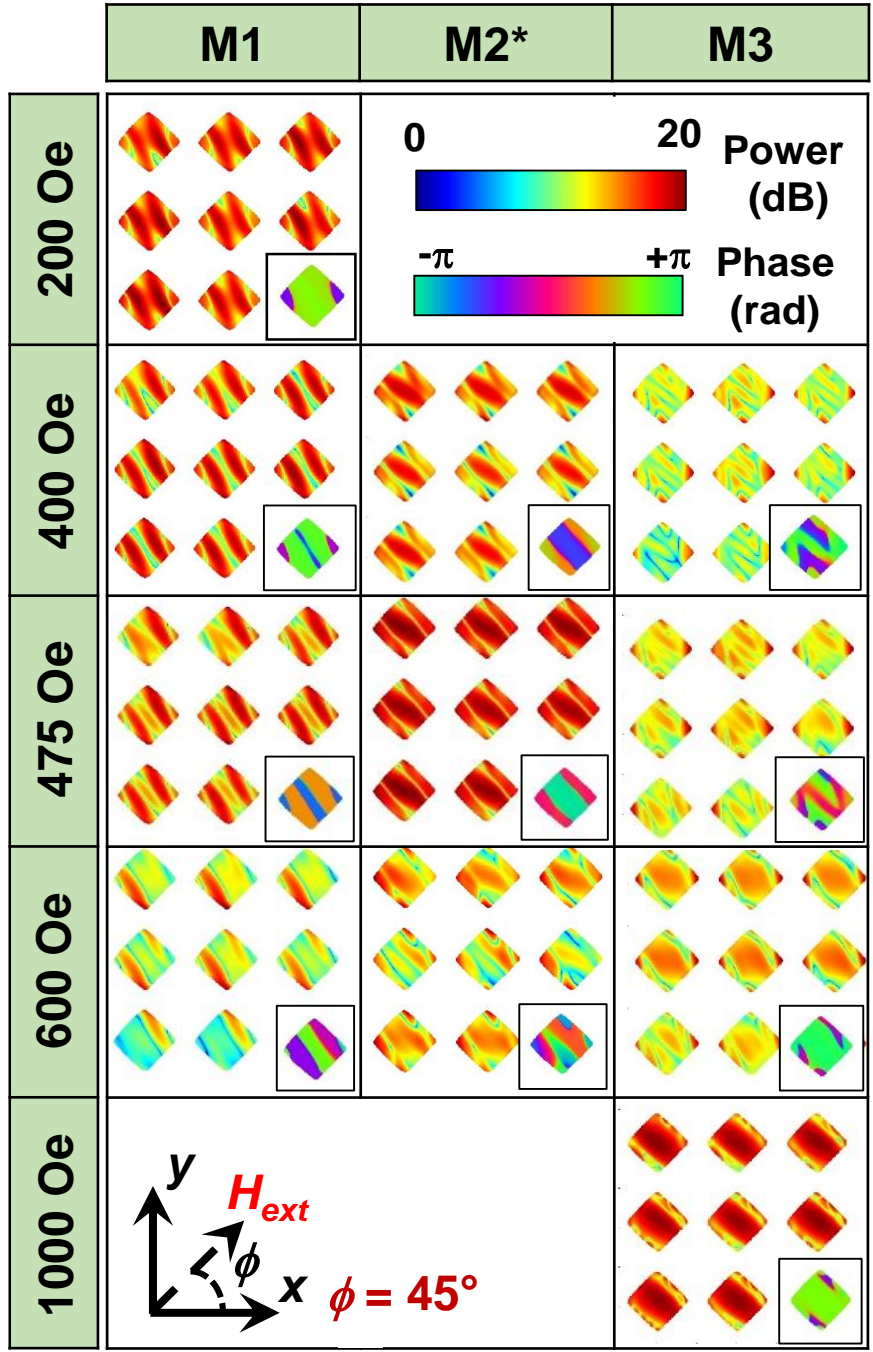

FIG. 6. Simulated spatial distribution of power and phase (in the inset) profiles corresponding to different SW modes at five different bias field values for $\phi = 45°$ for the Py nanodot array. The applied field direction is shown at the bottom left of the figure. Symbols with different colors represent different SW modes. The color map is shown at the upper right side of the figure.

### 3.2.3. Distribution of Exchange field

To understand the origin of the magnon-magnon coupling and its modulation with bias magnetic field, we have simulated the spatial distribution of the dipole-exchange field (Exchange field distribution of each dot, which is modulated by dipolar interaction of nanodot array) lines at the equilibrium for different bias field orientations. Fig. 7 shows the exchange field map of nanodots array at eight different fields for $\phi = 45°$ orientation. Due to inter-dot dipolar interactions, a dynamic variation of exchange field line with the bias field amplitude (for better viewing purpose, we just present a single nanodot) is observed. The Supplementary Movie A1 shows the dynamics of this exchange field in more detail. At lower bias fields ($H_{ext} \ll H_{ac}$), due to dominating effect of demagnetizing field, spins take a configuration such that at equilibrium condition the exchange field lines create three different regions within a single dot. The field lines of center and edge regions are configured in opposite direction as denoted with yellow and green arrows in Fig. 7(a). As we increase the bias field, the region around the edge of the dot start to vanish and the center region gradually expands. At a very high bias field ($H_{ext} \gg H_{ac}$), e.g., $H_{ext} = 1000$ Oe, only the central region with unidirectional field lines are observed inside a dot. This transformation from three mutually opposite (antiparallel) field-line configuration to uniform (parallel) configuration occurs for 450 Oe $\leq H_{ext} \leq$ 500 Oe, which is exactly the anticrossing field region for $\phi = 45°$ orientation. This change in exchange field profile can be observed much more clearly if we take a linescan along the bias field direction (white dotted line in Fig. 7(a)) as shown in Fig. 7(b). In the inset, we have magnified the end part of the linescan. Here, it is clearly visible that below the anticrossing field ($H_{ext} = H_{ac} = 475$ Oe) the linescan has two different local maxima which transform into one maximum as we increase $H_{ext}$. The exchange field profile for other values of $\phi$ are shown in section S7 of the Supplementary Materials, where similar transformation is observed in the anticrossing field region. Our observation of correlation between these two phenomena indicates that the anticrossing gap appears only when such a variation of exchange field occurs due to the bias field strength as well as its orientation. The internal field distribution in presence and absence of the exchange field leads to similar conclusion, which we have described in section S8 of the Supplementary Materials.

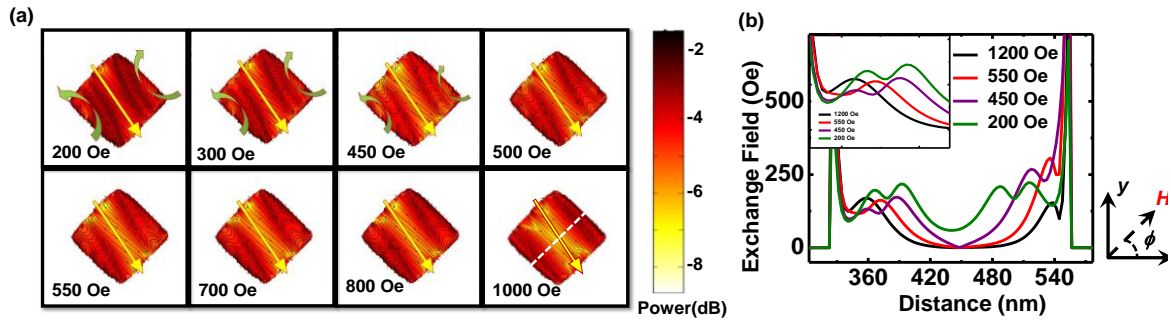

FIG. 7. Exchange field distributions for (a) single nanodot for eight different bias field values at $\phi = 45°$. Yellow and green arrows represent the direction of exchange field at the center and edge position of the nanodot. We have shown here a single nanodot from the center of the array for clarity in spin configurations. The color bars are shown at the right side of the figure. (b) Linescan of the simulated exchange field for nanodot array along the field direction. In the inset magnified portion of simulated exchange field is shown.

## 4. Conclusion

In summary, the interaction between magnons confined in a sole magnonic cavity has been realized in the strong coupling regime. We have investigated a bias field strength and angle-dependent magnetization dynamics in diamond-shaped Py nanodot arrays using the broadband ferromagnetic resonance technique. Our study has demonstrated that the coupling between two magnon modes is mediated by the exchange coupling inside individual nanodot. Furthermore, the coupling strength is found to be highly dependent on the orientation and strength of the bias magnetic field, leading towards the possibility of externally controlled hybrid magnonic devices. The experimental results have been well reproduced by micromagnetic simulation. The power and phase profiles of the resonant modes have been numerically calculated to gain insight into the spatial nature of the dynamics. The transformation of power as well as mode property from one branch of SW to another, apparently support the strong interaction in-between these modes. Numerical study shows that the anticrossing gap appears when the symmetry of exchange configuration inside each nanodot is broken due to the applied bias magnetic field. We have also observed mode softening phenomena when the static magnetic configuration switches from the S-state to the leaf state and with the variation of bias field angle it gradually disappears. Our findings offer a new approach toward tunable magnon-magnon coupling in ferromagnetic nanostructures for applications in quantum transduction using magnons.

## 5. Acknowledgements

AB gratefully acknowledges the financial support from S. N. Bose National Centre for Basic Sciences, India (Grant No. SNB/AB/18-19/211). SB acknowledges Science and Engineering Research Board (SERB), India for funding (Grant no. CRG/2018/002080). SM and SC acknowledge S. N. Bose National Centre for Basic Sciences for senior research fellowship